\documentclass[%
 aip,
 jmp,%
 amsmath,amssymb,
 reprint,%
]{revtex4-1}

\usepackage{graphicx}% Include figure files
\usepackage{dcolumn}% Align table columns on decimal point
\usepackage{bm}% bold math
\usepackage{subfigure}
\begin{document}

%\preprint{AIP/123-QED}

\title{Magnetoresistance in an electronic cavity coupled to one-dimensional systems }% Force line breaks with \\

\author{Chengyu Yan}
 \email{uceeya3@ucl.ac.uk}
\author{Sanjeev Kumar}%
\affiliation{%
	London Centre for Nanotechnology, 17-19 Gordon Street, London WC1H 0AH, United Kingdom\\
}%
\affiliation{
	Department of Electronic and Electrical Engineering, University College London, Torrington Place, London WC1E 7JE, United Kingdom
}%
\author{Patrick See}
\affiliation{%
	National Physical Laboratory, Hampton Road, Teddington, Middlesex TW11 0LW, United Kingdom\\
}%

\author{Ian Farrer}
\altaffiliation[Currently at ]{Department of Electronic and Electrical Engineering, University of Sheffield, Mappin Street, Sheffield S1 3JD, United Kingdom.}%Lines break automatically or can be forced with \\
\author{David Ritchie}
\author{J. P. Griffiths}
\author{G. A. C. Jones}
\affiliation{%
	Cavendish Laboratory, J.J. Thomson Avenue, Cambridge CB3 OHE, United Kingdom\\
}%
\author{Michael Pepper}
\affiliation{%
	London Centre for Nanotechnology, 17-19 Gordon Street, London WC1H 0AH, United Kingdom\\
}%
\affiliation{
	Department of Electronic and Electrical Engineering, University College London, Torrington Place, London WC1E 7JE, United Kingdom
}%

\date{\today}% It is always \today

\begin{abstract}

In the present work we performed magnetoresistance measurement in a hybrid system consisting of an arc-shaped quantum point contact (QPC) and a flat, rectangular QPC, both of which together form an electronic cavity between them. The results highlight a transition between collimation-induced resistance dip to a magnetoresistance peak as the strength of coupling between the QPC and the electronic cavity was increased. The initial results show the promise of hybrid quantum system for future quantum technologies.

\end{abstract}

\maketitle

Recent development in quantum technologies has stimulated research activities in integrating different quantum components in order to realize complex functionality\cite{HCR07,MCG07}. It is therefore of fundamental interest to investigate coupling between discrete quantum devices. Coupling between electronic cavity and other quantum devices, such as quantum point contact\cite{KEA97, DTW01,YKP16,YKP17,SPK17} (QPC) and quantum dot\cite{ROZ15,DLL17,FOR17} (QD), has attracted considerable attention. A hybrid device consisting of a QPC and an electronic cavity, as an example, provides a unique platform to investigate electronic equivalent of optical phenomena. This may be understood from the fact that electrons in such a system transport ballistically and accumulate phase along the quasi-classical trajectories, which is a close analogue of an optical cavity. Previous studies based on QPC-cavity hybrid devices reported results based on classical trajectories of electrons\cite{KEA97, DTW01,HHH99,HHH00} as well as quantum effects manifested as conductance fluctuations\cite{KEA97, DTW01} and Ahronov-Bohm phase shift as a function of cavity size\cite{DTW01}.

In the present work, we studied magnetoresistance in a hybrid system in a controlled manner  with the assistance of two QPCs which form an electronic cavity between them. We show the strength of coupling between the QPC and cavity states can be monitored by oscillation in the magnitude of central peak/dip in magnetoresistance.

\begin{figure}
	\includegraphics[height=2.0in,width=3.2in]{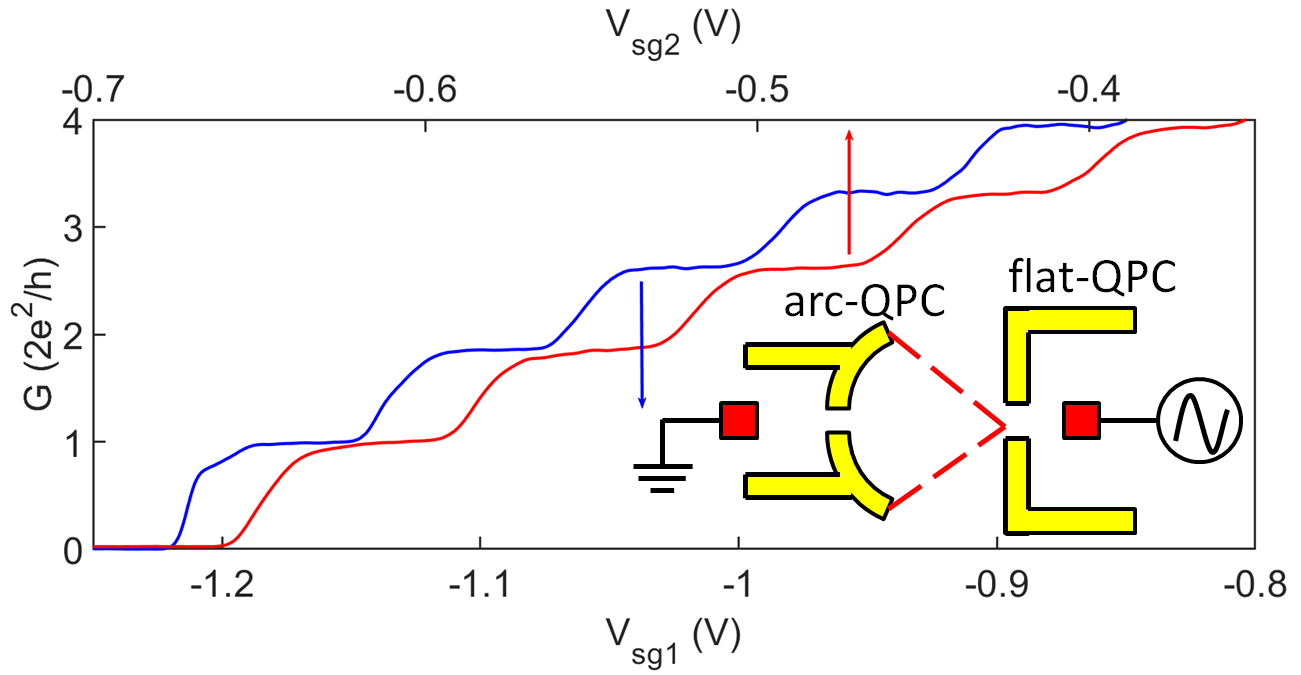} 
	
	\caption{The experiment setup and device characteristics. The blue trace shows the characteristic of arc-QPC as a function of gate voltage V$_{sg1}$; the red trace illustrates the behaviour of flat-QPC against gate voltage V$_{sg2}$. The series resistance was not removed. Inset depicts an illustration of the experiment setup, the yellow blocks represent electron-beam lithographically defined metallic gates while the red squares highlight the Ohmic contact. The length (width) of the flat-QPC is 700 nm (500 nm). The radius of the arc is 2 $\mu$m with an opening angle of 45$^\circ$. Both the length and width of the QPC formed in the center of the arc, i.e arc-QPC, are 200 nm.
	}           
	\label{fig:1}
\end{figure} 

\begin{figure*}
	\includegraphics[height=2.5in,width=6.0in]{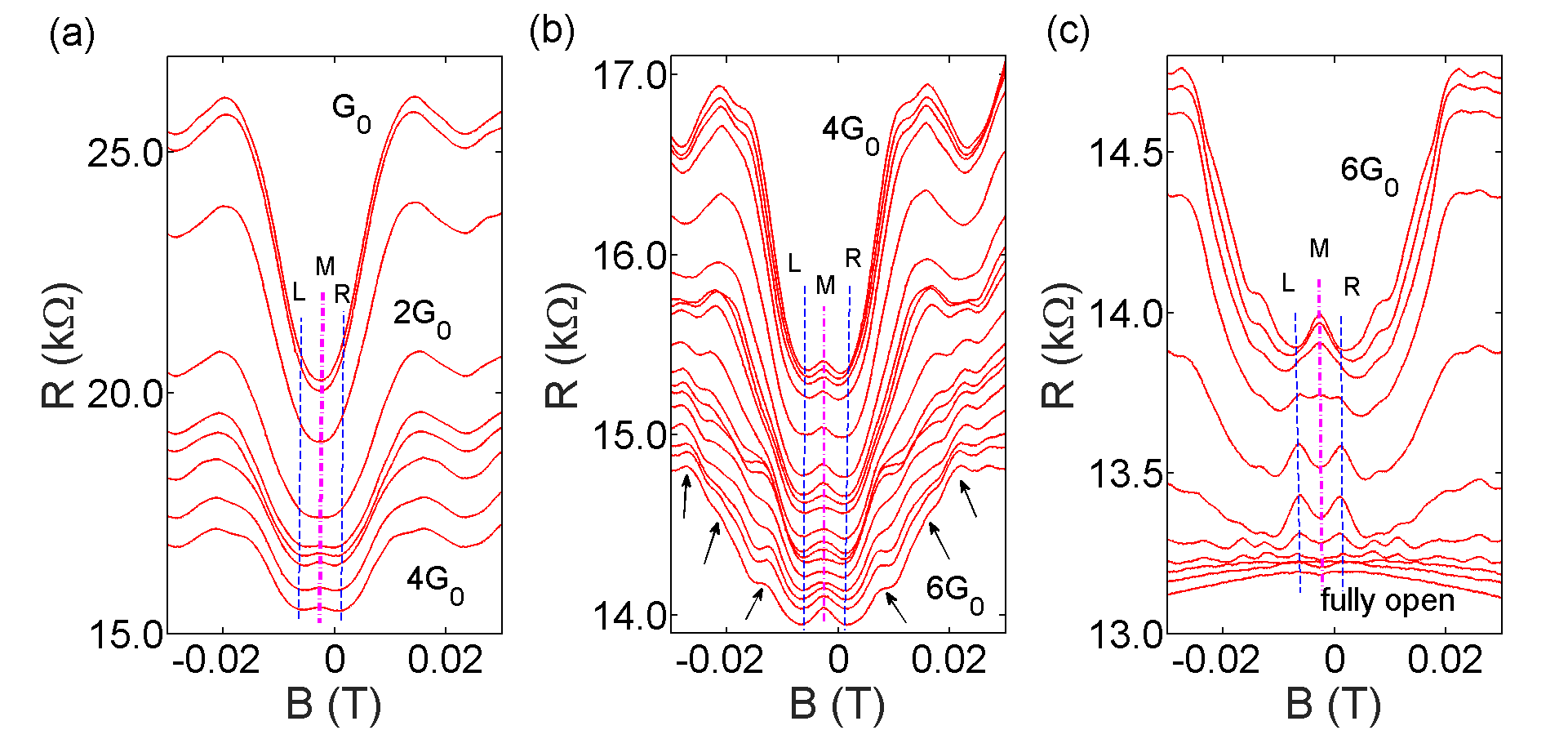}
	
	\caption{Magnetoresistance of the hybrid system with flat-QPC as an emitter. The flat-QPC conductance was incremented while arc-QPC was fixed at G$_0$ (G$_0$ = $\frac{2e^2}{h}$). (a) Result in regime 1 (G$_0$ to 4G$_0$), the central dip gradually evolved into a peak with increasing flat-QPC conductance. (b) Result in regime 2 (4G$_0$ to 6G$_0$), the central peak is present in this regime. The black arrows highlight the satellite peaks. (c) Result in regime 3 (6G$_0$ to channel fully open), the central peak split into two peaks in  the 1D-2D transition regime and eventually all features are smeared out. The strength of the central feature is defined as such $\Delta$R = R$_M$-(R$_L$+R$_R$)/2, where R$_M$, R$_L$ and R$_R$ refer to the resistance measured at given magnetic field marked by the vertical dashed lines.    
	}           
	\label{fig:2}
\end{figure*}

\begin{figure*}
	
	\includegraphics[height=2.5in,width=6.0in]{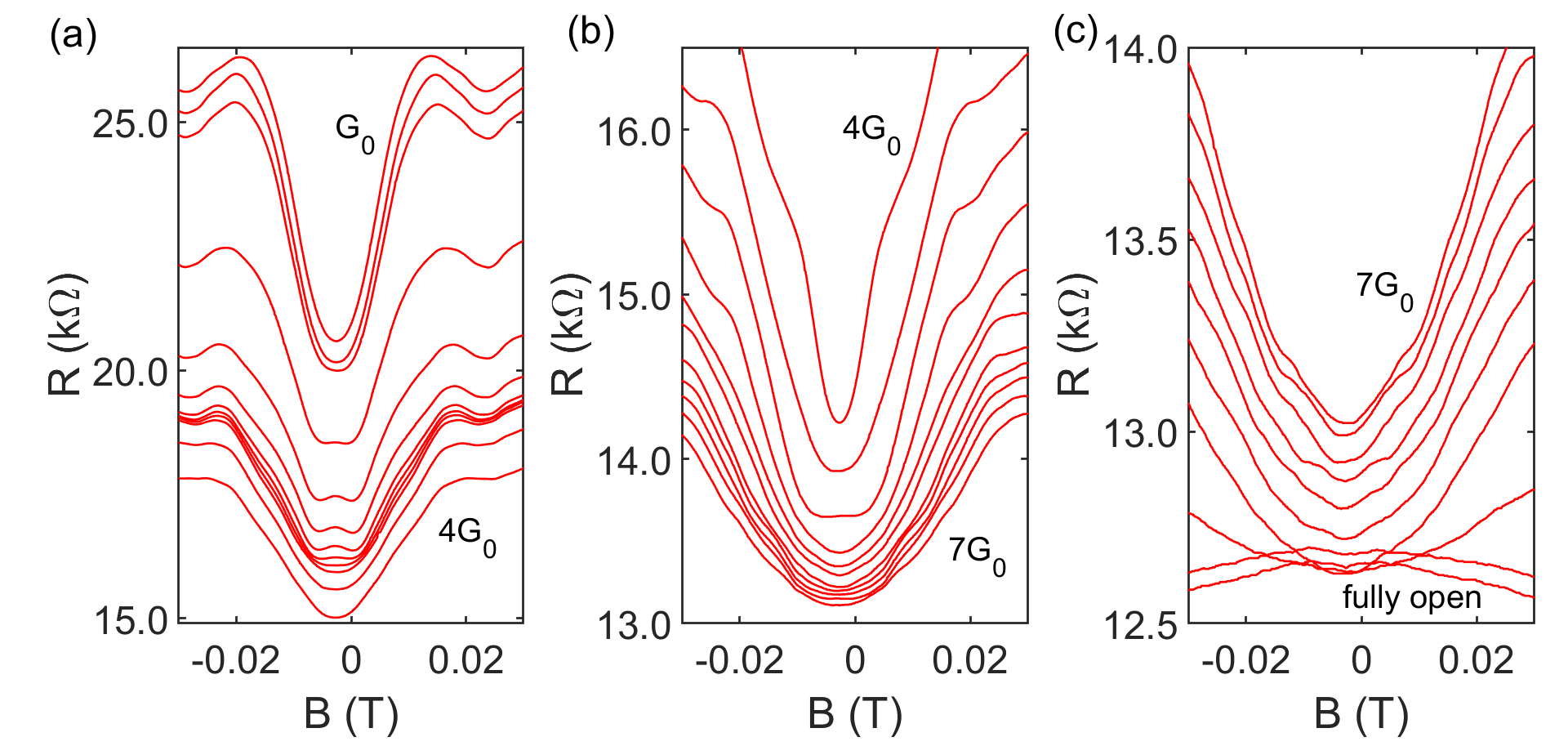}
	
	\caption{Magnetoresistance of the hybrid system with arc-QPC as an emitter. The arc-QPC conductance was incremented while flat-QPC was fixed at G$_0$. (a)-(c) show results in all the three regimes, regime 1 (G$_0$ to 4G$_0$), regime 2 (4G$_0$ to 7G$_0$), and regime 3 (7G$_0$-channel fully open), respectively. It was seen that the central dip dominated the spectrum.}           
	\label{fig:3}
	
\end{figure*} 

The devices studied in the work were fabricated from a high mobility two-dimensional electron gas (2DEG) formed at the interface of GaAs/Al$_{0.33}$Ga$_{0.67}$As heterostructure. The measured electron density (mobility) was 1.80$\times$10$^{11}$cm$^{-2}$ (2.17$\times$10$^6$cm$^2$V$^{-1}$s$^{-1}$) at 1.5 K, which ensured that both the calculated mean free path and phase coherence length \cite{AAK82,PPP89} were over 10 $\mu$m which were larger than electron propagation length. The experiments were performed in a cryofree dilution refrigerator with a lattice temperature of 20 mK using the standard lockin technique. 

The hybrid device consists of a pair of arc-shaped gates with a QPC (referred as arc-QPC) forming in the center of arc-gates and another pair of rectangular QPC (named as flat-QPC) as depicted in Fig.~\ref{fig:1}. The QPCs are assembled in such a way that the geometrical center of the arc (shaped gates) aligns with the saddle point of the flat-QPC. An electronic cavity is formed when QPCs are activated by depleting the 2D electrons underneath the gates\cite{YKP16,YKP17}. Both the arc-QPC and flat-QPC showed well defined one-dimensional conductance quantization when they were characterised individually, Fig.~\ref{fig:1}. 

In the presence of a small transverse magnetic field, the magnetoresistance of flat-QPC or arc-QPC exhibited a weak-localization peak similar to reported previously\cite{BHD95,KIL96}. However, the non-trivial features started appearing when the hybrid device was formed, i.e. both flat-QPC and arc-QPC were activated.

In the first experiment, the flat-QPC served as an emitter while the arc-QPC was used as a collector, see inset of Fig.~\ref{fig:1}. The voltage applied to the flat-QPC was incremented slowly corresponding to a conductance of G$_0$ (G$_0$=$\frac{2e^2}{h}$) up to 1D channel fully open while the arc-QPC was fixed at G$_0$. The magnetoresistance was investigated in three different regimes according to flat-QPC conductance. 
 
In regime 1, the flat-QPC was incremented from G$_0$ to 4G$_0$, Fig.~\ref{fig:2}(a). A dip in resistance (marked by the magenta dashed line) was observed around 0 T when the flat-QPC conductance G $\leqslant$ 2G$_0$ which is due to the fact that the injected electrons had a relatively small angular spread owing to strong collimation in low conductance regime\cite{LAC90, HBG98}. The electrons tend to propagate from the flat-QPC through the arc-QPC directly without backscattering; however, the applied magnetic field guides the injected electrons to the arc-shaped boundary wall of the arc-QPC and thus results in backscattering, which in turn triggers a rise in  resistance.  In this respect, our hybrid system is similar to a long quantum wire where scattering at the boundary was suggested to introduce a central dip in magnetoresistance\cite{TRS89}. An offset in central dip in magnetoresistance of 3 mT could be due to magnetic hysteresis of the superconducting magnet. On increasing G to 4G$_0$, a central magnetoresistance peak started forming. The zero-field magnetoresistance peak in electronic billiards is a result of geometry induced closed loop\cite{BKM94} (in other words, an analogue to weak localization). A large angular spread at higher G makes injected electrons to be reflected at the boundary wall of the arc-shaped QPCs, thus forming a close loop even at zero magnetic field; on the hand, a relatively small angular spread at low conductance makes such reflection unlikely to happen without the assistance of a magnetic field. The backscattered electrons will be refocused to the saddle point of flat-QPC. 
 
In regime 2, Fig.~\ref{fig:2}(b), the flat-QPC was set from 4G$_0$ to 6G$_0$, the magnitude of the central peak fluctuated  in the sense that the central peak gradually smeared out when the flat-QPC conductance was close to 5G$_0$, and then reappeared on further increasing the conductance of flat-QPC. 
The fluctuation will be discussed in detail in Fig.~\ref{fig:5}. Meanwhile it was also noticed that multiple weak-satellite peaks, marked by black arrows in Fig.~\ref{fig:2}(b), occurred in this regime. It was suggested\cite{DTW01} in a previous work that the appearance of these satellite peaks was an indication of Aharonov-Bohm effect and each peak was associated with a particular classical orbit. We suggest that although the satellite peaks might be relevant with classical orbits, however, Aharonov-Bohm effect did not occur in our experiment considering the fact that the satellites peaks were almost absent in regime 1 or regime 3.   
 
In regime 3 (6G$_0$ to fully open emitter), Fig.~\ref{fig:2}(c), the central peak gradually splits into two peaks around the 1D-2D transition regime of the flat-QPC and eventually all the features smeared out and only a smooth background was observed with the flat-QPC entering into the 2D regime. The smooth background agrees well the weak-localization signal when the arc-QPC was characterised individually.

To be noted that Shubnikov-de Haas oscillation started appearing in all the three regimes when the magnetic field exceeded $\pm$0.13 T (data is not shown). 

\begin{figure}
	
	\includegraphics[height=1.0in,width=3.6in]{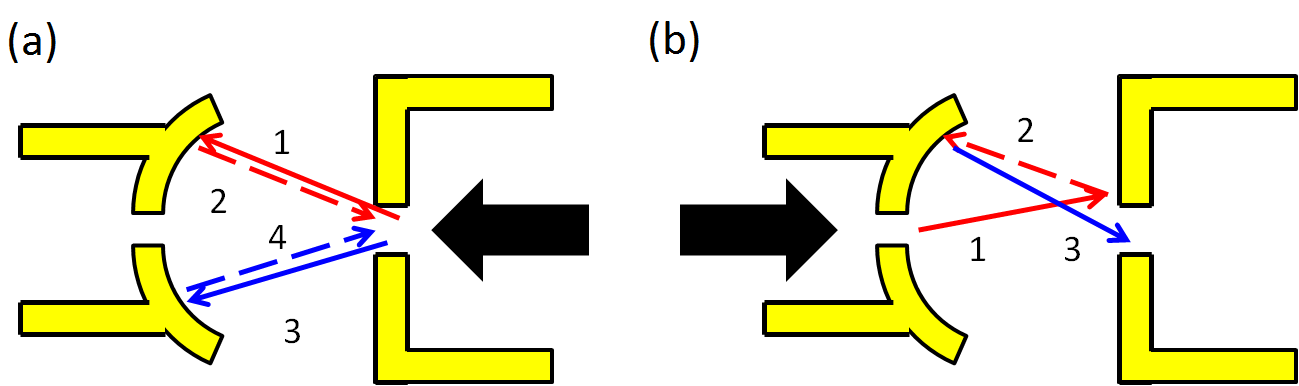}
	
	\caption{Representative electron trajectories with flat-QPC and arc-QPC acting as emitter, respectively. The solid traces represent the trajectory of incident electrons whereas the dashed traces illustrate the reflected electrons. In plot (a), the solid and dashed traces are offset intentionally for clarity, which otherwise should overlap together. The thick black arrows indicate current injection direction.
	}           
	\label{fig:4}
\end{figure}

To ensure the observation did not simply arise from the superposition of the magneto-spectrum of two individual QPCs, we reversed the role of emitter and collector. In setup II the arc-QPC was utilized as an emitter and incremented  while the flat-QPC functioned as a collector and was fixed at G$_0$. In addition, the ac signal is fed to the left Ohmic [Fig.~\ref{fig:1}(a)] whereas the right Ohmic is grounded in setup II.The results are summarized in Fig.~\ref{fig:3}. Results in regime 1, Fig.~\ref{fig:3}(a), was similar to that observed with setup I. However, the central dip dominated in regime 2 [Fig.~\ref{fig:3}(b)] and 3 [Fig.~\ref{fig:3}(c)] which was considerably different from its counterpart in Fig.~\ref{fig:2} where more features were resolved. The behaviour in setup II was similar to the magnetoresistance in two regular QPC in series\cite{HBG98}. It is interesting to mention that satellite peaks observed in  Fig.~\ref{fig:2}(b) did not occur in setup II. A comparison between setup I and II also suggests that the complicated evolution of magnetoresistance observed in Fig.~\ref{fig:2} did not directly arise from the form of wavefunction at different emitter conductance; otherwise, setup II should exhibit similar behaviour.

The difference between the results from two setup could be understood with a semi-classical picture as shown in Fig.~\ref{fig:4}. Electrons injected from the flat-QPC, which aligns with the geometrical centre of the arc (i.e. arc-QPC), experience an arc-shaped reflector which traps the electrons in an electronic cavity defined by these QPCs. The injected electrons after reflection at the boundary wall of the arc would be directed towards the flat-QPC. Owing to the geometry of cavity defined between the arc- and flat-QPCs, electrons would be trapped in a closed loop such as events 1$\rightarrow$4 as shown in Fig.~\ref{fig:4}(a) until the total propagation length exceeded the mean free path; phase associated with such a close loop is unlikely to be averaged out, therefore corrections to the resistance, i.e. the central magnetoresistance peak, due to the accumulated phase was observable. On the other hand, the trajectory of electrons injected from the arc-QPC, i.e. setup II, did not necessarily form a closed loop, so that it was relatively easy for the injected electrons to get through the hybrid system via a series of scattering events, for instance events 1$\rightarrow$3 as depicted in Fig.~\ref{fig:4}(b). Electron trajectory in the second scenario is more arbitrary and the trajectory-determined phase tends to be averaged out, which leads to no obvious corrections in the resistance.

\begin{figure}
	
	\includegraphics[height=2.0in,width=3.6in]{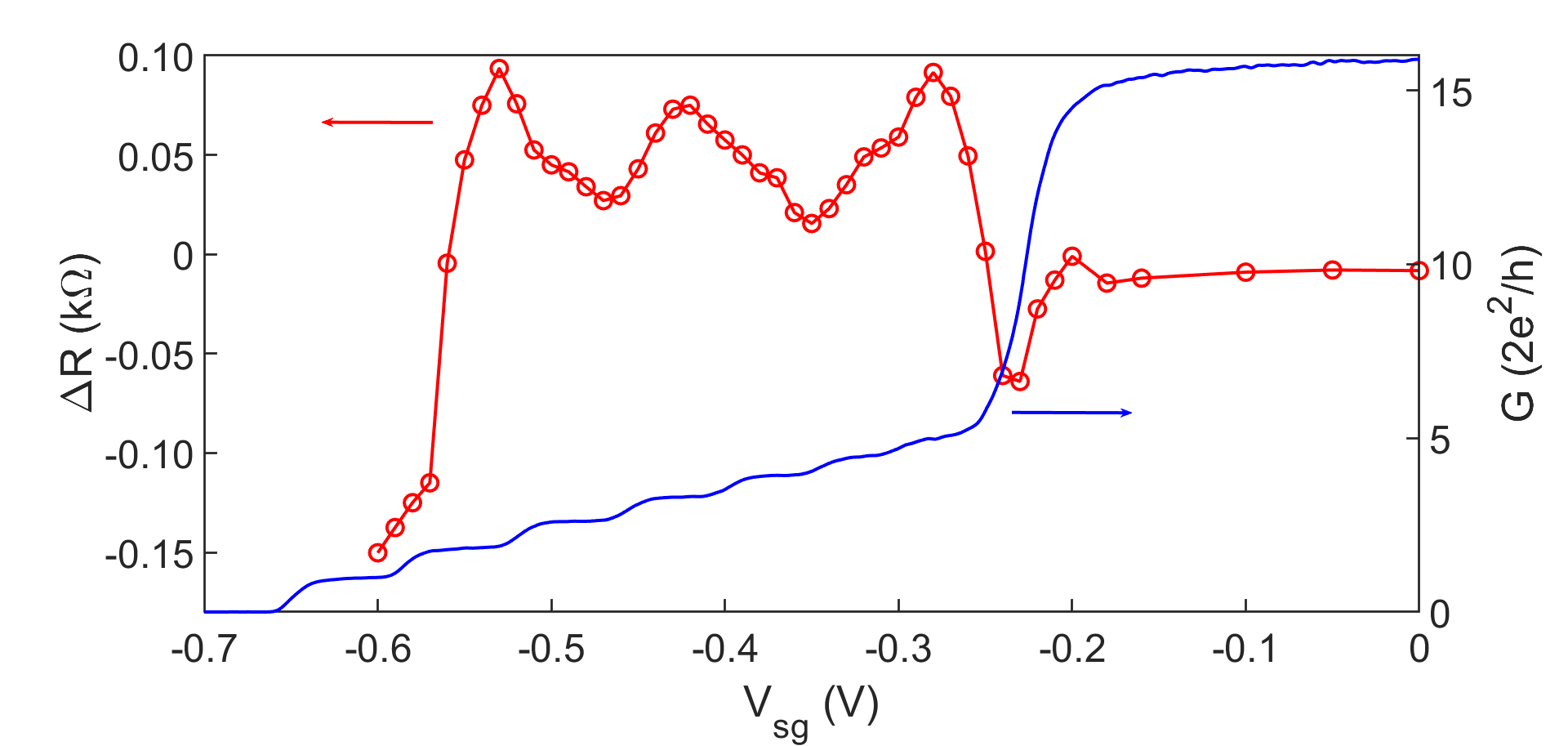}
	
	\caption{Fluctuation of the central feature as a function of flat-QPC conductance. The relative strength of the central feature, $\Delta$R = R$_M$-(R$_L$+R$_R$)/2, shows quasi-periodic oscillation, where R$_M$, R$_L$ and R$_R$ refer to resistance at the given magnetic field marked in Fig.~\ref{fig:2}.
	}           
	\label{fig:5}
\end{figure}

After addressing the difference between the two setup, we discuss a possible mechanism behind the observed fluctuation of the central features with flat-QPC serving as an emitter. To quantify the fluctuation, we defined the strength of the central feature (could be dip or peak) as such $\Delta$R = R$_M$-(R$_L$+R$_R$)/2, where R$_M$, R$_L$ and R$_R$ refer to the resistance measured at given magnetic field marked in Fig.~\ref{fig:2} (although in dip dominant regime there was not noticeble feature at $L$ or $R$, we still use the resistance at the same field for systematic investigation). It was seen that $\Delta$R followed a quasi-periodic oscillation\cite{SUO94,YKP16,YKP17,KEA97, DTW01} when the flat-QPC was tuned into the 1D regime (V$_{sg}$ $\leqslant$ -0.25 V); the fluctuation smeared out when the flat-QPC entered the 2D regime as shown in Fig. 5. The fact that the peak of oscillation does not necessarily occur at each conductance plateau suggesting that it is not simply associated with occupation of 1D subband or electron collimation, which would otherwise produce peaks corresponding to each conductance plateau. Instead, the oscillation was an indication of the coupling between the cavity and QPC sates. Each peak in Fig. 5 is a result of removing a cavity mode, therefore  peaks in $\Delta$R should occur when the change in radius $r$ of cavity matched a condition \cite{KEA97}, $\Delta r$ = N$\times$$\lambda_F$/2, where N is an integer and $\lambda_F$ is the Fermi wavelength.    

In conclusion we have shown magnetoresistance in a hybrid system consisting of QPCs coupled via an electronic cavity. It was found that the central magneto-feature around 0 T underwent a transition from dip into peak when the cavity was present whereas resistance dip dominated when the cavity was effectively absent. An oscillation of the strength of the central magneto-feature was observed as a consequence of coupling between the QPC and cavity sates. The results provide insight of coupling between discrete quantum devices which is valuable for further development of integrated quantum systems.    

The work is funded by the Engineering and Physical Sciences Research Council (EPSRC), United Kingdom.

%\bibliography{Mag_QPC_cav}% Produces the bibliography via BibTeX.

\begin{thebibliography}{21}%
		\makeatletter
		\providecommand \@ifxundefined [1]{%
			\@ifx{#1\undefined}
		}%
		\providecommand \@ifnum [1]{%
			\ifnum #1\expandafter \@firstoftwo
			\else \expandafter \@secondoftwo
			\fi
		}%
		\providecommand \@ifx [1]{%
			\ifx #1\expandafter \@firstoftwo
			\else \expandafter \@secondoftwo
			\fi
		}%
		\providecommand \natexlab [1]{#1}%
		\providecommand \enquote  [1]{``#1''}%
		\providecommand \bibnamefont  [1]{#1}%
		\providecommand \bibfnamefont [1]{#1}%
		\providecommand \citenamefont [1]{#1}%
		\providecommand \href@noop [0]{\@secondoftwo}%
		\providecommand \href [0]{\begingroup \@sanitize@url \@href}%
		\providecommand \@href[1]{\@@startlink{#1}\@@href}%
		\providecommand \@@href[1]{\endgroup#1\@@endlink}%
		\providecommand \@sanitize@url [0]{\catcode `\\12\catcode `\$12\catcode
			`\&12\catcode `\#12\catcode `\^12\catcode `\_12\catcode `\%12\relax}%
		\providecommand \@@startlink[1]{}%
		\providecommand \@@endlink[0]{}%
		\providecommand \url  [0]{\begingroup\@sanitize@url \@url }%
		\providecommand \@url [1]{\endgroup\@href {#1}{\urlprefix }}%
		\providecommand \urlprefix  [0]{URL }%
		\providecommand \Eprint [0]{\href }%
		\providecommand \doibase [0]{http://dx.doi.org/}%
		\providecommand \selectlanguage [0]{\@gobble}%
		\providecommand \bibinfo  [0]{\@secondoftwo}%
		\providecommand \bibfield  [0]{\@secondoftwo}%
		\providecommand \translation [1]{[#1]}%
		\providecommand \BibitemOpen [0]{}%
		\providecommand \bibitemStop [0]{}%
		\providecommand \bibitemNoStop [0]{.\EOS\space}%
		\providecommand \EOS [0]{\spacefactor3000\relax}%
		\providecommand \BibitemShut  [1]{\csname bibitem#1\endcsname}%
		\let\auto@bib@innerbib\@empty
		%</preamble>
		\bibitem [{\citenamefont {Hu}\ \emph {et~al.}(2007)\citenamefont {Hu},
			\citenamefont {Churchill}, \citenamefont {Reilly}, \citenamefont {Xiang},
			\citenamefont {Lieber},\ and\ \citenamefont {Marcus}}]{HCR07}%
		\BibitemOpen
		\bibfield  {author} {\bibinfo {author} {\bibfnamefont {Y.}~\bibnamefont
				{Hu}}, \bibinfo {author} {\bibfnamefont {H.~O.}\ \bibnamefont {Churchill}},
			\bibinfo {author} {\bibfnamefont {D.~J.}\ \bibnamefont {Reilly}}, \bibinfo
			{author} {\bibfnamefont {J.}~\bibnamefont {Xiang}}, \bibinfo {author}
			{\bibfnamefont {C.~M.}\ \bibnamefont {Lieber}}, \ and\ \bibinfo {author}
			{\bibfnamefont {C.~M.}\ \bibnamefont {Marcus}},\ }\bibfield  {title}
		{\enquote {\bibinfo {title} {A Ge/Si heterostructure nanowire-based double
					quantum dot with integrated charge sensor},}\ }\href@noop {} {\bibfield
			{journal} {\bibinfo  {journal} {Nature nanotechnology}\ }\textbf {\bibinfo
				{volume} {2}},\ \bibinfo {pages} {622--625} (\bibinfo {year}
			{2007})}\BibitemShut {NoStop}%
		\bibitem [{\citenamefont {Majer}\ \emph {et~al.}(2007)\citenamefont {Majer},
			\citenamefont {Chow}, \citenamefont {Gambetta}, \citenamefont {Koch},
			\citenamefont {Johnson}, \citenamefont {Schreier}, \citenamefont {Frunzio},
			\citenamefont {Schuster}, \citenamefont {Houck}, \citenamefont {Wallraff}
			\emph {et~al.}}]{MCG07}%
		\BibitemOpen
		\bibfield  {author} {\bibinfo {author} {\bibfnamefont {J.}~\bibnamefont
				{Majer}}, \bibinfo {author} {\bibfnamefont {J.}~\bibnamefont {Chow}},
			\bibinfo {author} {\bibfnamefont {J.}~\bibnamefont {Gambetta}}, \bibinfo
			{author} {\bibfnamefont {J.}~\bibnamefont {Koch}}, \bibinfo {author}
			{\bibfnamefont {B.}~\bibnamefont {Johnson}}, \bibinfo {author} {\bibfnamefont
				{J.}~\bibnamefont {Schreier}}, \bibinfo {author} {\bibfnamefont
				{L.}~\bibnamefont {Frunzio}}, \bibinfo {author} {\bibfnamefont
				{D.}~\bibnamefont {Schuster}}, \bibinfo {author} {\bibfnamefont
				{A.}~\bibnamefont {Houck}}, \bibinfo {author} {\bibfnamefont
				{A.}~\bibnamefont {Wallraff}},  \emph {et~al.},\ }\bibfield  {title}
		{\enquote {\bibinfo {title} {Coupling superconducting qubits via a cavity
					bus.}}\ }\href@noop {} {\bibfield  {journal} {\bibinfo  {journal} {Nature}\
			}\textbf {\bibinfo {volume} {449}},\ \bibinfo {pages} {443--447} (\bibinfo
			{year} {2007})}\BibitemShut {NoStop}%
		\bibitem [{\citenamefont {Katine}\ \emph {et~al.}(1997)\citenamefont {Katine},
			\citenamefont {Eriksson}, \citenamefont {Adourian}, \citenamefont
			{Westervelt}, \citenamefont {Edwards}, \citenamefont {Lupu-Sax},
			\citenamefont {Heller}, \citenamefont {Campman},\ and\ \citenamefont
			{Gossard}}]{KEA97}%
		\BibitemOpen
		\bibfield  {author} {\bibinfo {author} {\bibfnamefont {J.~A.}\ \bibnamefont
				{Katine}}, \bibinfo {author} {\bibfnamefont {M.~A.}\ \bibnamefont
				{Eriksson}}, \bibinfo {author} {\bibfnamefont {A.~S.}\ \bibnamefont
				{Adourian}}, \bibinfo {author} {\bibfnamefont {R.~M.}\ \bibnamefont
				{Westervelt}}, \bibinfo {author} {\bibfnamefont {J.~D.}\ \bibnamefont
				{Edwards}}, \bibinfo {author} {\bibfnamefont {A.}~\bibnamefont {Lupu-Sax}},
			\bibinfo {author} {\bibfnamefont {E.~J.}\ \bibnamefont {Heller}}, \bibinfo
			{author} {\bibfnamefont {K.~L.}\ \bibnamefont {Campman}}, \ and\ \bibinfo
			{author} {\bibfnamefont {A.~C.}\ \bibnamefont {Gossard}},\ }\bibfield
		{title} {\enquote {\bibinfo {title} {Point contact conductance of an open
					resonator},}\ }\href {\doibase 10.1103/PhysRevLett.79.4806} {\bibfield
			{journal} {\bibinfo  {journal} {Phys. Rev. Lett.}\ }\textbf {\bibinfo
				{volume} {79}},\ \bibinfo {pages} {4806--4809} (\bibinfo {year}
			{1997})}\BibitemShut {NoStop}%
		\bibitem [{\citenamefont {Duncan}\ \emph {et~al.}(2001)\citenamefont {Duncan},
			\citenamefont {Topinka}, \citenamefont {Westervelt}, \citenamefont
			{Maranowski},\ and\ \citenamefont {Gossard}}]{DTW01}%
		\BibitemOpen
		\bibfield  {author} {\bibinfo {author} {\bibfnamefont {D.~S.}\ \bibnamefont
				{Duncan}}, \bibinfo {author} {\bibfnamefont {M.~A.}\ \bibnamefont {Topinka}},
			\bibinfo {author} {\bibfnamefont {R.~M.}\ \bibnamefont {Westervelt}},
			\bibinfo {author} {\bibfnamefont {K.~D.}\ \bibnamefont {Maranowski}}, \ and\
			\bibinfo {author} {\bibfnamefont {A.~C.}\ \bibnamefont {Gossard}},\
		}\bibfield  {title} {\enquote {\bibinfo {title} {Aharonov-bohm phase shift in
					an open electron resonator},}\ }\href {\doibase 10.1103/PhysRevB.64.033310}
		{\bibfield  {journal} {\bibinfo  {journal} {Phys. Rev. B}\ }\textbf {\bibinfo
				{volume} {64}},\ \bibinfo {pages} {033310} (\bibinfo {year}
			{2001})}\BibitemShut {NoStop}%
		\bibitem [{\citenamefont {Yan}\ \emph {et~al.}(2017{\natexlab{a}})\citenamefont
			{Yan}, \citenamefont {Kumar}, \citenamefont {Pepper}, \citenamefont {See},
			\citenamefont {Farrer}, \citenamefont {Ritchie}, \citenamefont {Griffiths},\
			and\ \citenamefont {Jones}}]{YKP16}%
		\BibitemOpen
		\bibfield  {author} {\bibinfo {author} {\bibfnamefont {C.}~\bibnamefont
				{Yan}}, \bibinfo {author} {\bibfnamefont {S.}~\bibnamefont {Kumar}}, \bibinfo
			{author} {\bibfnamefont {M.}~\bibnamefont {Pepper}}, \bibinfo {author}
			{\bibfnamefont {P.}~\bibnamefont {See}}, \bibinfo {author} {\bibfnamefont
				{I.}~\bibnamefont {Farrer}}, \bibinfo {author} {\bibfnamefont
				{D.}~\bibnamefont {Ritchie}}, \bibinfo {author} {\bibfnamefont
				{J.}~\bibnamefont {Griffiths}}, \ and\ \bibinfo {author} {\bibfnamefont
				{G.}~\bibnamefont {Jones}},\ }\bibfield  {title} {\enquote {\bibinfo {title}
				{Fano resonance in a cavity-reflector hybrid system},}\ }\href {\doibase
			10.1103/PhysRevB.95.041407} {\bibfield  {journal} {\bibinfo  {journal} {Phys.
					Rev. B}\ }\textbf {\bibinfo {volume} {95}},\ \bibinfo {pages} {041407}
			(\bibinfo {year} {2017}{\natexlab{a}})}\BibitemShut {NoStop}%
		\bibitem [{\citenamefont {Yan}\ \emph {et~al.}(2017{\natexlab{b}})\citenamefont
			{Yan}, \citenamefont {Kumar}, \citenamefont {Pepper}, \citenamefont {See},
			\citenamefont {Farrer}, \citenamefont {Ritchie}, \citenamefont {Griffiths},\
			and\ \citenamefont {Jones}}]{YKP17}%
		\BibitemOpen
		\bibfield  {author} {\bibinfo {author} {\bibfnamefont {C.}~\bibnamefont
				{Yan}}, \bibinfo {author} {\bibfnamefont {S.}~\bibnamefont {Kumar}}, \bibinfo
			{author} {\bibfnamefont {M.}~\bibnamefont {Pepper}}, \bibinfo {author}
			{\bibfnamefont {P.}~\bibnamefont {See}}, \bibinfo {author} {\bibfnamefont
				{I.}~\bibnamefont {Farrer}}, \bibinfo {author} {\bibfnamefont
				{D.}~\bibnamefont {Ritchie}}, \bibinfo {author} {\bibfnamefont
				{J.}~\bibnamefont {Griffiths}}, \ and\ \bibinfo {author} {\bibfnamefont
				{G.}~\bibnamefont {Jones}},\ }\bibfield  {title} {\enquote {\bibinfo {title}
				{Interference effects in a tunable quantum point contact integrated with an
					electronic cavity},}\ }\href {\doibase 10.1103/PhysRevApplied.8.024009}
		{\bibfield  {journal} {\bibinfo  {journal} {Phys. Rev. Applied}\ }\textbf
			{\bibinfo {volume} {8}},\ \bibinfo {pages} {024009} (\bibinfo {year}
			{2017}{\natexlab{b}})}\BibitemShut {NoStop}%
		\bibitem [{\citenamefont {Steinacher}\ \emph {et~al.}(2017)\citenamefont
			{Steinacher}, \citenamefont {P{\"o}ltl}, \citenamefont {Kr{\"a}henmann},
			\citenamefont {Hofmann}, \citenamefont {Reichl}, \citenamefont {Zwerger},
			\citenamefont {Wegscheider}, \citenamefont {Jalabert}, \citenamefont
			{Ensslin}, \citenamefont {Weinmann} \emph {et~al.}}]{SPK17}%
		\BibitemOpen
		\bibfield  {author} {\bibinfo {author} {\bibfnamefont {R.}~\bibnamefont
				{Steinacher}}, \bibinfo {author} {\bibfnamefont {C.}~\bibnamefont
				{P{\"o}ltl}}, \bibinfo {author} {\bibfnamefont {T.}~\bibnamefont
				{Kr{\"a}henmann}}, \bibinfo {author} {\bibfnamefont {A.}~\bibnamefont
				{Hofmann}}, \bibinfo {author} {\bibfnamefont {C.}~\bibnamefont {Reichl}},
			\bibinfo {author} {\bibfnamefont {W.}~\bibnamefont {Zwerger}}, \bibinfo
			{author} {\bibfnamefont {W.}~\bibnamefont {Wegscheider}}, \bibinfo {author}
			{\bibfnamefont {R.}~\bibnamefont {Jalabert}}, \bibinfo {author}
			{\bibfnamefont {K.}~\bibnamefont {Ensslin}}, \bibinfo {author} {\bibfnamefont
				{D.}~\bibnamefont {Weinmann}},  \emph {et~al.},\ }\bibfield  {title}
		{\enquote {\bibinfo {title} {Scanning gate experiments: from strongly to
					weakly invasive probes},}\ }\href@noop {} {\bibfield  {journal} {\bibinfo
				{journal} {arXiv preprint arXiv:1709.08559}\ } (\bibinfo {year}
			{2017})}\BibitemShut {NoStop}%
		\bibitem [{\citenamefont {R\"ossler}\ \emph {et~al.}(2015)\citenamefont
			{R\"ossler}, \citenamefont {Oehri}, \citenamefont {Zilberberg}, \citenamefont
			{Blatter}, \citenamefont {Karalic}, \citenamefont {Pijnenburg}, \citenamefont
			{Hofmann}, \citenamefont {Ihn}, \citenamefont {Ensslin}, \citenamefont
			{Reichl},\ and\ \citenamefont {Wegscheider}}]{ROZ15}%
		\BibitemOpen
		\bibfield  {author} {\bibinfo {author} {\bibfnamefont {C.}~\bibnamefont
				{R\"ossler}}, \bibinfo {author} {\bibfnamefont {D.}~\bibnamefont {Oehri}},
			\bibinfo {author} {\bibfnamefont {O.}~\bibnamefont {Zilberberg}}, \bibinfo
			{author} {\bibfnamefont {G.}~\bibnamefont {Blatter}}, \bibinfo {author}
			{\bibfnamefont {M.}~\bibnamefont {Karalic}}, \bibinfo {author} {\bibfnamefont
				{J.}~\bibnamefont {Pijnenburg}}, \bibinfo {author} {\bibfnamefont
				{A.}~\bibnamefont {Hofmann}}, \bibinfo {author} {\bibfnamefont
				{T.}~\bibnamefont {Ihn}}, \bibinfo {author} {\bibfnamefont {K.}~\bibnamefont
				{Ensslin}}, \bibinfo {author} {\bibfnamefont {C.}~\bibnamefont {Reichl}}, \
			and\ \bibinfo {author} {\bibfnamefont {W.}~\bibnamefont {Wegscheider}},\
		}\bibfield  {title} {\enquote {\bibinfo {title} {Transport spectroscopy of a
					spin-coherent dot-cavity system},}\ }\href {\doibase
			10.1103/PhysRevLett.115.166603} {\bibfield  {journal} {\bibinfo  {journal}
				{Phys. Rev. Lett.}\ }\textbf {\bibinfo {volume} {115}},\ \bibinfo {pages}
			{166603} (\bibinfo {year} {2015})}\BibitemShut {NoStop}%
		\bibitem [{\citenamefont {Dias~da Silva}\ \emph {et~al.}(2017)\citenamefont
			{Dias~da Silva}, \citenamefont {Lewenkopf}, \citenamefont {Vernek},
			\citenamefont {Ferreira},\ and\ \citenamefont {Ulloa}}]{DLL17}%
		\BibitemOpen
		\bibfield  {author} {\bibinfo {author} {\bibfnamefont {L.~G. G.~V.}\
				\bibnamefont {Dias~da Silva}}, \bibinfo {author} {\bibfnamefont {C.~H.}\
				\bibnamefont {Lewenkopf}}, \bibinfo {author} {\bibfnamefont {E.}~\bibnamefont
				{Vernek}}, \bibinfo {author} {\bibfnamefont {G.~J.}\ \bibnamefont
				{Ferreira}}, \ and\ \bibinfo {author} {\bibfnamefont {S.~E.}\ \bibnamefont
				{Ulloa}},\ }\bibfield  {title} {\enquote {\bibinfo {title} {Conductance and
					kondo interference beyond proportional coupling},}\ }\href {\doibase
			10.1103/PhysRevLett.119.116801} {\bibfield  {journal} {\bibinfo  {journal}
				{Phys. Rev. Lett.}\ }\textbf {\bibinfo {volume} {119}},\ \bibinfo {pages}
			{116801} (\bibinfo {year} {2017})}\BibitemShut {NoStop}%
		\bibitem [{\citenamefont {Ferguson}\ \emph {et~al.}(2017)\citenamefont
			{Ferguson}, \citenamefont {Oehri}, \citenamefont {R\"ossler}, \citenamefont
			{Ihn}, \citenamefont {Ensslin}, \citenamefont {Blatter},\ and\ \citenamefont
			{Zilberberg}}]{FOR17}%
		\BibitemOpen
		\bibfield  {author} {\bibinfo {author} {\bibfnamefont {M.~S.}\ \bibnamefont
				{Ferguson}}, \bibinfo {author} {\bibfnamefont {D.}~\bibnamefont {Oehri}},
			\bibinfo {author} {\bibfnamefont {C.}~\bibnamefont {R\"ossler}}, \bibinfo
			{author} {\bibfnamefont {T.}~\bibnamefont {Ihn}}, \bibinfo {author}
			{\bibfnamefont {K.}~\bibnamefont {Ensslin}}, \bibinfo {author} {\bibfnamefont
				{G.}~\bibnamefont {Blatter}}, \ and\ \bibinfo {author} {\bibfnamefont
				{O.}~\bibnamefont {Zilberberg}},\ }\bibfield  {title} {\enquote {\bibinfo
				{title} {Long-range spin coherence in a strongly coupled all-electronic
					dot-cavity system},}\ }\href {\doibase 10.1103/PhysRevB.96.235431} {\bibfield
			{journal} {\bibinfo  {journal} {Phys. Rev. B}\ }\textbf {\bibinfo {volume}
				{96}},\ \bibinfo {pages} {235431} (\bibinfo {year} {2017})}\BibitemShut
		{NoStop}%
		\bibitem [{\citenamefont {Hersch}, \citenamefont {Haggerty},\ and\
			\citenamefont {Heller}(1999)}]{HHH99}%
		\BibitemOpen
		\bibfield  {author} {\bibinfo {author} {\bibfnamefont {J.~S.}\ \bibnamefont
				{Hersch}}, \bibinfo {author} {\bibfnamefont {M.~R.}\ \bibnamefont
				{Haggerty}}, \ and\ \bibinfo {author} {\bibfnamefont {E.~J.}\ \bibnamefont
				{Heller}},\ }\bibfield  {title} {\enquote {\bibinfo {title} {Diffractive
					orbits in an open microwave billiard},}\ }\href {\doibase
			10.1103/PhysRevLett.83.5342} {\bibfield  {journal} {\bibinfo  {journal}
				{Phys. Rev. Lett.}\ }\textbf {\bibinfo {volume} {83}},\ \bibinfo {pages}
			{5342--5345} (\bibinfo {year} {1999})}\BibitemShut {NoStop}%
		\bibitem [{\citenamefont {Hersch}, \citenamefont {Haggerty},\ and\
			\citenamefont {Heller}(2000)}]{HHH00}%
		\BibitemOpen
		\bibfield  {author} {\bibinfo {author} {\bibfnamefont {J.~S.}\ \bibnamefont
				{Hersch}}, \bibinfo {author} {\bibfnamefont {M.~R.}\ \bibnamefont
				{Haggerty}}, \ and\ \bibinfo {author} {\bibfnamefont {E.~J.}\ \bibnamefont
				{Heller}},\ }\bibfield  {title} {\enquote {\bibinfo {title} {Influence of
					diffraction on the spectrum and wave functions of an open system},}\ }\href
		{\doibase 10.1103/PhysRevE.62.4873} {\bibfield  {journal} {\bibinfo
				{journal} {Phys. Rev. E}\ }\textbf {\bibinfo {volume} {62}},\ \bibinfo
			{pages} {4873--4888} (\bibinfo {year} {2000})}\BibitemShut {NoStop}%
		\bibitem [{\citenamefont {Altshuler}, \citenamefont {Aronov},\ and\
			\citenamefont {Khmelnitsky}(1982)}]{AAK82}%
		\BibitemOpen
		\bibfield  {author} {\bibinfo {author} {\bibfnamefont {B.~L.}\ \bibnamefont
				{Altshuler}}, \bibinfo {author} {\bibfnamefont {A.}~\bibnamefont {Aronov}}, \
			and\ \bibinfo {author} {\bibfnamefont {D.}~\bibnamefont {Khmelnitsky}},\
		}\bibfield  {title} {\enquote {\bibinfo {title} {Effects of electron-electron
					collisions with small energy transfers on quantum localisation},}\
		}\href@noop {} {\bibfield  {journal} {\bibinfo  {journal} {Journal of Physics
					C: Solid State Physics}\ }\textbf {\bibinfo {volume} {15}},\ \bibinfo {pages}
			{7367} (\bibinfo {year} {1982})}\BibitemShut {NoStop}%
		\bibitem [{\citenamefont {Pooke}\ \emph {et~al.}(1989)\citenamefont {Pooke},
			\citenamefont {Paquin}, \citenamefont {Pepper},\ and\ \citenamefont
			{Gundlach}}]{PPP89}%
		\BibitemOpen
		\bibfield  {author} {\bibinfo {author} {\bibfnamefont {D.~M.}\ \bibnamefont
				{Pooke}}, \bibinfo {author} {\bibfnamefont {N.}~\bibnamefont {Paquin}},
			\bibinfo {author} {\bibfnamefont {M.}~\bibnamefont {Pepper}}, \ and\ \bibinfo
			{author} {\bibfnamefont {A.}~\bibnamefont {Gundlach}},\ }\bibfield  {title}
		{\enquote {\bibinfo {title} {Electron-electron scattering in narrow Si
					accumulation layers},}\ }\href@noop {} {\bibfield  {journal} {\bibinfo
				{journal} {Journal of Physics: Condensed Matter}\ }\textbf {\bibinfo {volume}
				{1}},\ \bibinfo {pages} {3289} (\bibinfo {year} {1989})}\BibitemShut
		{NoStop}%
		\bibitem [{\citenamefont {Reulet}, \citenamefont {Bouchiat},\ and\
			\citenamefont {Mailly}(1995)}]{BHD95}%
		\BibitemOpen
		\bibfield  {author} {\bibinfo {author} {\bibfnamefont {B.}~\bibnamefont
				{Reulet}}, \bibinfo {author} {\bibfnamefont {H.}~\bibnamefont {Bouchiat}}, \
			and\ \bibinfo {author} {\bibfnamefont {D.}~\bibnamefont {Mailly}},\
		}\bibfield  {title} {\enquote {\bibinfo {title} {Magnetoconductance, weak
					localization and electron-electron interactions in semi-ballistic quantum
					wires},}\ }\href@noop {} {\bibfield  {journal} {\bibinfo  {journal} {EPL
					(Europhysics Letters)}\ }\textbf {\bibinfo {volume} {31}},\ \bibinfo {pages}
			{305} (\bibinfo {year} {1995})}\BibitemShut {NoStop}%
		\bibitem [{\citenamefont {Koester}\ \emph {et~al.}(1996)\citenamefont
			{Koester}, \citenamefont {Ismail}, \citenamefont {Lee},\ and\ \citenamefont
			{Chu}}]{KIL96}%
		\BibitemOpen
		\bibfield  {author} {\bibinfo {author} {\bibfnamefont {S.~J.}\ \bibnamefont
				{Koester}}, \bibinfo {author} {\bibfnamefont {K.}~\bibnamefont {Ismail}},
			\bibinfo {author} {\bibfnamefont {K.~Y.}\ \bibnamefont {Lee}}, \ and\
			\bibinfo {author} {\bibfnamefont {J.~O.}\ \bibnamefont {Chu}},\ }\bibfield
		{title} {\enquote {\bibinfo {title} {Weak localization in back-gated
					Si/${\mathrm{Si}}_{0.7}$${\mathrm{Ge}}_{0.3}$ quantum-well wires fabricated
					by reactive ion etching},}\ }\href {\doibase 10.1103/PhysRevB.54.10604}
		{\bibfield  {journal} {\bibinfo  {journal} {Phys. Rev. B}\ }\textbf {\bibinfo
				{volume} {54}},\ \bibinfo {pages} {10604--10608} (\bibinfo {year}
			{1996})}\BibitemShut {NoStop}%
		\bibitem [{\citenamefont {Molenkamp}\ \emph {et~al.}(1990)\citenamefont
			{Molenkamp}, \citenamefont {Staring}, \citenamefont {Beenakker},
			\citenamefont {Eppenga}, \citenamefont {Timmering}, \citenamefont
			{Williamson}, \citenamefont {Harmans},\ and\ \citenamefont {Foxon}}]{LAC90}%
		\BibitemOpen
		\bibfield  {author} {\bibinfo {author} {\bibfnamefont {L.~W.}\ \bibnamefont
				{Molenkamp}}, \bibinfo {author} {\bibfnamefont {A.~A.~M.}\ \bibnamefont
				{Staring}}, \bibinfo {author} {\bibfnamefont {C.~W.~J.}\ \bibnamefont
				{Beenakker}}, \bibinfo {author} {\bibfnamefont {R.}~\bibnamefont {Eppenga}},
			\bibinfo {author} {\bibfnamefont {C.~E.}\ \bibnamefont {Timmering}}, \bibinfo
			{author} {\bibfnamefont {J.~G.}\ \bibnamefont {Williamson}}, \bibinfo
			{author} {\bibfnamefont {C.~J. P.~M.}\ \bibnamefont {Harmans}}, \ and\
			\bibinfo {author} {\bibfnamefont {C.~T.}\ \bibnamefont {Foxon}},\ }\bibfield
		{title} {\enquote {\bibinfo {title} {Electron-beam collimation with a quantum
					point contact},}\ }\href {\doibase 10.1103/PhysRevB.41.1274} {\bibfield
			{journal} {\bibinfo  {journal} {Phys. Rev. B}\ }\textbf {\bibinfo {volume}
				{41}},\ \bibinfo {pages} {1274--1277} (\bibinfo {year} {1990})}\BibitemShut
		{NoStop}%
		\bibitem [{\citenamefont {Heindrichs}\ \emph {et~al.}(1998)\citenamefont
			{Heindrichs}, \citenamefont {Buhmann}, \citenamefont {Godijn},\ and\
			\citenamefont {Molenkamp}}]{HBG98}%
		\BibitemOpen
		\bibfield  {author} {\bibinfo {author} {\bibfnamefont {A.~S.~D.}\
				\bibnamefont {Heindrichs}}, \bibinfo {author} {\bibfnamefont
				{H.}~\bibnamefont {Buhmann}}, \bibinfo {author} {\bibfnamefont {S.~F.}\
				\bibnamefont {Godijn}}, \ and\ \bibinfo {author} {\bibfnamefont {L.~W.}\
				\bibnamefont {Molenkamp}},\ }\bibfield  {title} {\enquote {\bibinfo {title}
				{Classical rebound trajectories in nonlocal ballistic electron transport},}\
		}\href {\doibase 10.1103/PhysRevB.57.3961} {\bibfield  {journal} {\bibinfo
				{journal} {Phys. Rev. B}\ }\textbf {\bibinfo {volume} {57}},\ \bibinfo
			{pages} {3961--3965} (\bibinfo {year} {1998})}\BibitemShut {NoStop}%
		\bibitem [{\citenamefont {Thornton}\ \emph {et~al.}(1989)\citenamefont
			{Thornton}, \citenamefont {Roukes}, \citenamefont {Scherer},\ and\
			\citenamefont {Van~de Gaag}}]{TRS89}%
		\BibitemOpen
		\bibfield  {author} {\bibinfo {author} {\bibfnamefont {T.~J.}\ \bibnamefont
				{Thornton}}, \bibinfo {author} {\bibfnamefont {M.~L.}\ \bibnamefont
				{Roukes}}, \bibinfo {author} {\bibfnamefont {A.}~\bibnamefont {Scherer}}, \
			and\ \bibinfo {author} {\bibfnamefont {B.~P.}\ \bibnamefont {Van~de Gaag}},\
		}\bibfield  {title} {\enquote {\bibinfo {title} {Boundary scattering in
					quantum wires},}\ }\href {\doibase 10.1103/PhysRevLett.63.2128} {\bibfield
			{journal} {\bibinfo  {journal} {Phys. Rev. Lett.}\ }\textbf {\bibinfo
				{volume} {63}},\ \bibinfo {pages} {2128--2131} (\bibinfo {year}
			{1989})}\BibitemShut {NoStop}%
		\bibitem [{\citenamefont {Berry}\ \emph {et~al.}(1994)\citenamefont {Berry},
			\citenamefont {Katine}, \citenamefont {Marcus}, \citenamefont {Westervelt},\
			and\ \citenamefont {Gossard}}]{BKM94}%
		\BibitemOpen
		\bibfield  {author} {\bibinfo {author} {\bibfnamefont {M.}~\bibnamefont
				{Berry}}, \bibinfo {author} {\bibfnamefont {J.}~\bibnamefont {Katine}},
			\bibinfo {author} {\bibfnamefont {C.}~\bibnamefont {Marcus}}, \bibinfo
			{author} {\bibfnamefont {R.}~\bibnamefont {Westervelt}}, \ and\ \bibinfo
			{author} {\bibfnamefont {A.}~\bibnamefont {Gossard}},\ }\bibfield  {title}
		{\enquote {\bibinfo {title} {Weak localization and conductance fluctuations
					in a chaotic quantum dot},}\ }\href@noop {} {\bibfield  {journal} {\bibinfo
				{journal} {Surface science}\ }\textbf {\bibinfo {volume} {305}},\ \bibinfo
			{pages} {495--500} (\bibinfo {year} {1994})}\BibitemShut {NoStop}%
		\bibitem [{\citenamefont {Saito}\ \emph {et~al.}(1994)\citenamefont {Saito},
			\citenamefont {Usuki}, \citenamefont {Okada}, \citenamefont {Futatsugi},
			\citenamefont {Kiehl},\ and\ \citenamefont {Yokoyama}}]{SUO94}%
		\BibitemOpen
		\bibfield  {author} {\bibinfo {author} {\bibfnamefont {M.}~\bibnamefont
				{Saito}}, \bibinfo {author} {\bibfnamefont {T.}~\bibnamefont {Usuki}},
			\bibinfo {author} {\bibfnamefont {M.}~\bibnamefont {Okada}}, \bibinfo
			{author} {\bibfnamefont {T.}~\bibnamefont {Futatsugi}}, \bibinfo {author}
			{\bibfnamefont {R.}~\bibnamefont {Kiehl}}, \ and\ \bibinfo {author}
			{\bibfnamefont {N.}~\bibnamefont {Yokoyama}},\ }\bibfield  {title} {\enquote
			{\bibinfo {title} {Coupling between one-dimensional states in a quantum point
					contact and an electron waveguide},}\ }\href@noop {} {\bibfield  {journal}
			{\bibinfo  {journal} {Applied physics letters}\ }\textbf {\bibinfo {volume}
				{65}},\ \bibinfo {pages} {3087--3089} (\bibinfo {year} {1994})}\BibitemShut
		{NoStop}%
\end{thebibliography}

%

\end{document}